\documentclass[aps,twocolumn,pra,superscriptaddress,amsmath,amssymb]{revtex4-1} %% for PRB
\usepackage{dcolumn}
\usepackage{bm}
\usepackage{amsmath}
\usepackage{txfonts}
\usepackage[T1]{fontenc}
\usepackage{xspace}
\usepackage{ulem}
\usepackage{comment}
\newcommand{\kets}[1]{\lvert#1\rangle}

\newcommand{\ket}[1]{\left|#1\right>}

\newcommand{\image}[4]{
\begin{figure}[htb]
\begin{center}
\includegraphics[width=#1\columnwidth,clip]{#2}
\caption{
#3
}
\label{fig:#4}
\end{center}
\end{figure}
}

\newcommand{\equ}[2]{
\begin{equation}
#1
\label{eq:#2}
\end{equation}
}

\ifx\pdfoutput\undefined
\usepackage[dvipdfmx]{graphicx}
\usepackage[dvipdfmx]{hyperref}
\usepackage[dvipdfmx]{color}
\else
\usepackage{graphicx}
\usepackage{hyperref}
\usepackage{color}
\fi
\begin{document}
\let\emph\textit

\title{
%  Low-Energy Effective Model of a Spin-$S$ Anisotropic Kitaev Model\\
  % Topological nature of spins in a Spin-$S$ Kitaev model in the anisotropic limit
  Quantum and classical behavior of spin-$S$ Kitaev models in an anisotropic limit
}
\author{Tetsuya Minakawa}
\author{Joji Nasu}
\author{Akihisa Koga }
\affiliation{
  Department of Physics, Tokyo Institute of Technology,
  Meguro, Tokyo 152- 8551, Japan
}

\date{\today}
\begin{abstract}
  We study low-energy properties of spin-$S$ Kitaev models
  in an anisotropic limit.
  The effective form of a local conserved quantity is derived
  in the low-energy subspace.
  We find this is the same as that of $S=1/2$ case for the half-integer spins
  but shows a different form for the integer spins.
  Applying the perturbation theory to the anisotropic Kitaev model,
  we obtain the effective Hamiltonian.
  In the integer spin case,
  the effective model is equivalent to a free spin model
  under an uniform magnetic field,
  where quantum fluctuations are quenched.
  On the other hand, in the half-integer case,
  the system is described by the toric code Hamiltonian,
  where quantum fluctuations play a crucial role in the ground state.
  The boundary effect in the anisotropic Kitaev system is also discussed.
\end{abstract}
\maketitle

%%%%%%%%%%%%%%%%%%%%%%%%%%%%%%%%%%%%%%%%%
\section{Introduction}
%%%%%%%%%%%%%%%%%%%%%%%%%%%%%%%%%%%%%%%%%
Quantum spin liquid (QSL),
which does not exhibit any magnetic orders even at zero temperature,
has attracted much interests in condensed matter physics~\cite{ANDERSON1973153,ISI:000275366100033,Savary2017}.
One of typical examples is a ground state of the one-dimensional
quantum Heisenberg spin systems.
In the simple one-dimensional spin chains,
the existence of the gap reflects the topological nature dependent on the spin magnitude~\cite{PhysRevLett.50.1153,Haldane}; low-energy excitations are gapless for the half-integer spins,
while gapful for the integer spins.
In fact, corresponding low-energy excitations have been observed
in real materials such as $\rm CuCl_2\cdot 2N(C_5D_5)$ ($S=1/2$)~\cite{Endoh}
and Ni(C$_2$H$_8$N$_2$)$_2$NO$_2$(ClO$_4$) ($S=1$)~\cite{NENP1,NENP2}.
It is also known that the topology of lattice structure in addition to that associated with the spin magnitude affects ground-state properties in the one-dimensional
systems~\cite{Aringa,Sierra,Koga2002,Koga_1998,Fukui1,Fukui2}.

Another candidate for the QSL state has recently attracted considerable attention
in the two-dimensional quantum spin model, so-called the Kitaev model~\cite{KITAEV20062,hermanns2018physics,trebst2017kitaev}.
%an exactly solvable model suggested by Kitaev
% has attracted considerable attention
% as another example exhibiting the QSL ground-state in two dimensions.
This model describes the $S=1/2$ quantum spin system
with bond-dependent Ising interactions on a honeycomb lattice [see Fig.~\ref{fig:honeycomb}(a)]. Its ground state is exactly shown to be a QSL state,
where spin degrees of freedom are fractionalized into itinerant Majorana fermions and $Z_2$ fluxes.
In the anisotropic interaction limit of this model, it was also suggested that topological computation with anyons can be implemented.
While the Kitaev model was originally introduced as a simple model in the viewpoint of the quantum information,
Jackeli and Khaliullin~\cite{PhysRevLett.102.017205} suggested that this model provides a good description of magnetic interactions in certain insulating magnets with strong spin-orbit couplings.
A lot of works have been done intensively in candidate materials of this model such as $A_2$IrO${}_3$($A=$ Na, Li)~\cite{PhysRevB.82.064412,PhysRevLett.108.127203,PhysRevLett.109.266406,PhysRevLett.108.127204,Kitagawa2018nature,PhysRevLett.113.107201,PhysRevB.89.045117,PhysRevB.90.205126,PhysRevLett.114.077202,modic2014realization} and $\alpha$-RuCl${}_3$~\cite{PhysRevB.92.235119,PhysRevB.90.041112,PhysRevB.91.094422,PhysRevB.91.144420,PhysRevB.91.180401,PhysRevLett.114.147201,PhysRevB.96.041405,Baek2017,Sears2017,Wang2017,Zheng2017,janvsa2018observation}.
To understand how the Kitaev physics appears in these real materials,
additional interaction effects~\cite{PhysRevLett.105.027204,PhysRevLett.110.097204,PhysRevLett.112.077204,PhysRevB.87.064508,PhysRevLett.113.107201,Nakauchi,Tomishige} and
finite temperature, dynamical, and transport properties
have been studied so far~\cite{PhysRevLett.120.217205,PhysRevLett.113.197205,PhysRevB.93.174425,PhysRevLett.119.127204,PhysRevB.92.115122,PhysRevLett.117.157203,Kasahara2018,banerjee2018excitations,Nasu2016nphys,Banerjee2017,hirobe2017,PhysRevB.96.241107,Leahy2017,Liu2018,PhysRevLett.120.067202,Hentrich2018,Jiang2011,yadav2016kitaev,Zhu2018,PhysRevB.98.014418,PhysRevLett.117.277202,PhysRevB.98.060404,PhysRevB.98.060405,PhysRevB.98.060416,PhysRevB.98.054433,PhysRevB.98.054432,PhysRevB.96.134408}.
Moreover, other possibilities to generate the Kitaev coupling have been suggested
beyond the Jackeli-Khaliullin setup~\cite{PhysRevB.97.014408,PhysRevB.97.014407,Jang2018pre}.
In contrast to the quantum spin chains,
it is still unclear whether or not the qualitative difference between
half-integer and integer spins appears
in ground-state properties of the generalized Kitaev models with spin magnitude $S$.
% although it never appears in high temperature properties~\cite{oitmaa2018incipient}.
This is because the Kitaev models with $S>1/2$ cases are difficult to address the magnetic properties due to the absence of the exact solvability in the isotropic  case~\cite{PhysRevB.78.115116,suzuki2018thermal,JPSJ.87.063703,oitmaa2018incipient,rousochatzakis2018quantum}.

In this paper, we investigate spin-$S$ Kitaev models in the anisotropic limit,
where the interactions in one of three kinds of bonds
are much larger than the others.
This anisotropic model has an advantage to study ground-state properties
in the thermodynamic limit correctly,
in contrast to the isotropic Kitaev model.
Applying the perturbation expansion to the anisotropic models,
we derive the exactly solvable low-energy effective Hamiltonian,
where the qualitative difference in spins plays a crucial role
in ground-state properties.
In the half-integer spin cases, we obtain the toric code Hamiltonian
in the $8S$th order perturbation, which is essentially the same as the $S=1/2$ case.
On the other hand, the integer spin system is simply represented by
isolated spins under the magnetic field, whose Hamiltonian is
obtained by the $4S$th order perturbation.
The effect of the open boundary is also addressed.

This paper is organized as follows.
In Sec.~\ref{sec:model}, we introduce the spin-$S$ Kitaev models and their local conserved quantities.
In Sec.~\ref{sec:effect-spin-repr}, we show the explicit representation of
the local conserved quantity in the low-energy subspace of the anisotropic limit.
We derive the effective Hamiltonians,
applying the perturbation theory to the spin-$S$ Kitaev models in Sec.~\ref{sec:effective-model}.
%The boundary effect is discussed in Sec.~\ref{sec:boundary-effect}.
The summary is provided in the last section.

%%%%%%%%%%%%%%%%%%%%%%%%%%%%%%%%%%%%%%%%%%%%%%
\section{Model}\label{sec:model}
%%%%%%%%%%%%%%%%%%%%%%%%%%%%%%%%%%%%%%%%%%%%%%
We consider the Kitaev model on a honeycomb lattice,
which is given by the following Hamiltonian, as
\equ{{\cal H} = -J_x\sum_{\langle i,j \rangle_x}S_i^x S_j^x-J_y\sum_{\langle i,j \rangle_y}S_i^y S_j^y-J_z\sum_{\langle i,j \rangle_z}S_i^z S_j^z,}{hamiltonian}
where $S_i^\alpha$ is the $\alpha(=x,y,z)$ component of a spin-$S$ operator
at the $i$th site.
$J_\gamma$ is the exchange constant on the $\gamma(=x, y, z)$ bonds,
which connect the nearest neighbor sites $\langle i,j \rangle_\gamma$.
%%%%%%%%%%%%%%%%%%%%%%%%%%%%%%%%%%%%%%%%%%%%%%%
\image{1}
      {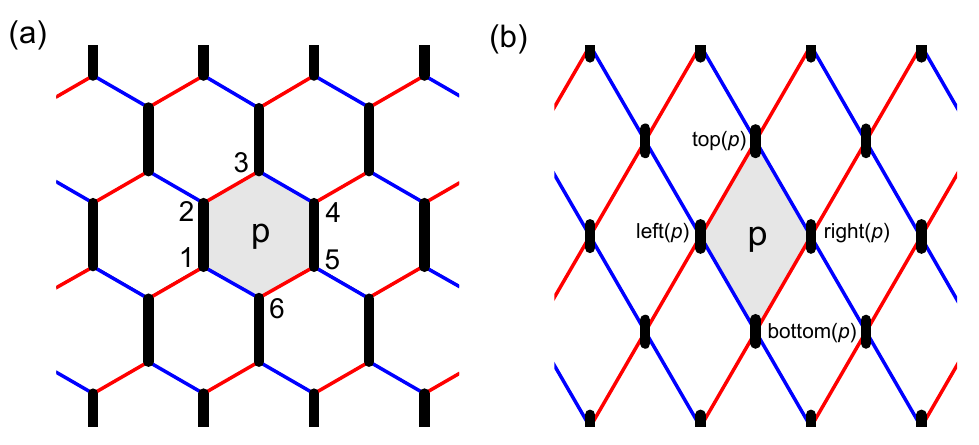}
      {(a) Lattice structure of the Kitaev model.
        $x$- and $y$-bonds are expressed by the thin red and blue lines and
        $z$-bonds by the bold lines.
        (b) Lattice structure of the effective model in the large $J_z$ limit.
        Each site has the pseudospin $\tilde{\sigma}$.
The shaded areas in (a) and (b) represent the fluxes $W_p$ and $W_p^{\rm eff}$ in the original and low-energy effective Hilbert spaces, respectively.
      }{honeycomb}
%%%%%%%%%%%%%%%%%%%%%%%%%%%%%%%%%%%%%%%%%%%%%%%
In Fig.~\ref{fig:honeycomb}(a), the Kitaev model is schematically depicted.
The exact solvability is lost in the general $S$ cases,
in contrast to the $S=1/2$ case.
However, even in the general spin cases, there also exists a local conserved quantity $W_p$ for each plaquette $p$,
which is given by~\cite{PhysRevB.78.115116}
\equ{W_p = \mp \exp\left[-i\pi (S_1^x + S_2 ^y + S_3^z + S_4^x + S_5^y + S_6^z)\right],}{cons}
where the sign $\mp$ is chosen to be $-$ for half-integer $S$ and $+$ for integer $S$, and the site positions $1, 2, \cdots, 6$ on the plaquette $p$ are indicated in Fig.~\ref{fig:honeycomb}(a).
From Eq.~(\ref{eq:cons}), $W_p^2=1$ is satisfied, and hence $W_p$ is a $Z_2$ conserved quantity.

In this study, we consider low-energy properties of the Kitaev model in the anisotropic limit,
$|J_x|, |J_y| \ll J_z$.
To this end, we split the Hamiltonian Eq.~(\ref{eq:hamiltonian}) into two parts;
${\cal H}_0=-J_z\sum_{\langle i,j \rangle_z}S_i^z S_j^z$ and
$V=-J_x\sum_{\langle i,j \rangle_x}S_i^x S_j^x-J_y\sum_{\langle i,j \rangle_y}S_i^y S_j^y$.
The ground states for ${\cal H}_0$ are $2^{N/2}$-fold degenerate with $E_0=-J_zS^2N/2$,
where $N$ is the total number of sites.
The subspace of ${\cal H}_0$ belonging to the ground state manifold is represented by direct products of
the fully polarized states
$\ket{+S}_i\ket{+S}_j$ or $\ket{-S}_i\ket{-S}_j$ at the dimer consisting of sites $i$ and $j$ on the $z$-bond,
where $\ket{m}_i$ is the local eigenstate of $S_i^z$ with the eigenvalue
$m(=-S,-S+1,\cdots,S)$ at site $i$.
This allows us to introduce the pseudospin $\tilde{\bm{\sigma}}$ on each $z$-bond so that
$\tilde{\sigma}^z_r|\tilde{\sigma}\rangle_r =\tilde{\sigma}|\tilde{\sigma}\rangle_r$ with $|\tilde{\uparrow}\rangle_r=|S\rangle_i|S\rangle_j$ and
$|\tilde{\downarrow}\rangle_r=|-S\rangle_i|-S\rangle_j$,
where $\tilde{\sigma}$ takes $+1$ ($\tilde{\uparrow}$) or $-1$ ($\tilde{\downarrow}$)
and $r$ specifies the location of the $z$-bond.

The ground-state degeneracy should be lifted by
introducing the perturbed Hamiltonian $V$.
In the conventional $S=1/2$ Kitaev model, the low-energy effective Hamiltonian is derived from the fourth order perturbation expansion with respect to $J_x$ and $J_y$~\cite{KITAEV20062}:
%%%%%%%%%%%%%%%%%%%%%%%%%%%%%%%%%%%%%%%%%
\begin{eqnarray}
  {\cal H}_{\rm eff}^{S=1/2} &=& -\frac{J_x^2J_y^2}{64J_z^3}\sum_p \tilde{\sigma}_{{\rm left}(p)}^y\tilde{\sigma}_{{\rm top}(p)}^z\tilde{\sigma}_{{\rm right}(p)}^y\tilde{\sigma}_{{\rm bottom}(p)}^z
\\
  &=&-\frac{J_x^2J_y^2}{64J_z^3}\sum_p W_p^{\rm eff},
  \label{eq:S=1/2}
\end{eqnarray}
%%%%%%%%%%%%%%%%%%%%%%%%%%%%%%%%%%%%%%%%%
where
%$r={\rm left}(p)$, ${\rm top}(p)$, ${\rm right}(p)$, and ${\rm bottom}(p)$ stand for the pseudospin positions on the plaquette $p$ shown in Fig.~\ref{fig:honeycomb}(b).
%We does not explicitly show a constant term in Eq.~(\ref{eq:S=1/2}) and also in the effective Hamiltonian obtained later.
we have used the fact that the local conserved quantity is represented by
$W_p^{\rm eff}= \tilde{\sigma}_{{\rm left}(p)}^y\tilde{\sigma}_{{\rm top}(p)}^z\tilde{\sigma}_{{\rm right}(p)}^y\tilde{\sigma}_{{\rm bottom}(p)}^z$ [see Fig.~\ref{fig:honeycomb}(b)].
This effective model is equivalent to the toric code Hamiltonian
under a unitary transformation associated with suitable spin
rotations~\cite{KITAEV20032}.
It is known that the unusual anyon excitations
appear due to quantum fluctuations.

In the Kitaev models generalized to spin $S>1/2$ cases,
the effective Hamiltonian is also represented by the pseudospin operators, which are introduced above.
However, it is nontrivial whether or not, in the anisotropic limit,
the generalized Kitaev models are described by the local conserved quantities
and are reduced to the toric code Hamiltonian.
In the following sections, we consider the local conserved quantities
and effective Hamiltonians in the anisotropic limit
to study the role of the spin magnitude in the system.

%%%%%%%%%%%%%%%%%%%%%%%%%%%%%%%%%%%%%%%%%%%%%%%%%%%%%%%%%
\section{effective-spin representation of local conserved quantities}
\label{sec:effect-spin-repr}
%%%%%%%%%%%%%%%%%%%%%%%%%%%%%%%%%%%%%%%%%%%%%%%%%%%%%%%%%

First, we focus on the local conserved quantity $W_p$,
which may provide a clue to understand the effective Hamiltonian of
the generalized Kitaev model in the anisotropic limit.
In the low-energy subspace,
neighboring spins on each $z$-bond are parallel and
fully polarized along the $z$ direction.
Therefore, the low-energy spin state on six sites of the plaquette $p$
[see Fig.~\ref{fig:honeycomb}(a)] should be specified as
$\ket{m_1}_1 \ket{m_1}_2 \ket{m_3}_3 \ket{m_4}_4 \ket{m_4}_5 \ket{m_6}_6$,
where $m_1,m_3,m_4$, and $m_6$ take $+S$ or $-S$.

To identify the effective form of $W_p$ in the low-energy subspace, we introduce
% Following equations for
 the rotations
by the angle $\pi$ about the $x$, $y$, and $z$ axis
% are convenient,
as
\begin{eqnarray}
%  \exp \left[-i \pi S^x\right]\ket{m}   &=&(-i)^{2S} \ket{-m},\\
%  \exp \left[-i \pi S^y\right]\ket{m}   &=&(-i)^{2m-2S} \ket{-m},\\
%  \exp \left[-i \pi S^z\right]\ket{m}   &=&(-i)^{2m} \ket{m}.
  \exp \left[-i \pi S^x\right]\ket{m}   &=&e^{-i\pi S} \ket{-m},\\
  \exp \left[-i \pi S^y\right]\ket{m}   &=&e^{-i\pi (m-S)} \ket{-m},\label{eq:y}\\
  \exp \left[-i \pi S^z\right]\ket{m}   &=&e^{-i\pi m} \ket{m}.\label{eq:z}
\end{eqnarray}
%%%%%%%%%%%%%%%%%%%%%%%%%%%%%%%%%%%%%%%%%%%%%%%%%%%%%%%%%
\begin{comment}
In general, the relative phase of the localized spin-$S$ states between $\kets{m}$ and $\kets{-m}$ is imposed as
{\color{red}
\equ{\exp \left[-i \pi S_j^x\right]\ket{m}_j   = \ket{-m}_j.}{rotationx}
}
Using the following relation:
\equ{\exp\left[ -i\pi S_j^y\right] = \exp \left[ -i \frac{\pi}{2} S_j^z\right] \exp\left[ -i \pi S_j^x\right] \exp\left[i \frac{\pi}{2} S_j^z\right],}{rotation2}
one can obtain the rotation around the $S^y$ axis
\equ{\exp\left[-i\pi S_j^y\right]\ket{m}_j = i^{2m}\ket{-m}_j.}{t2}
\end{comment}
%%%%%%%%%%%%%%%%%%%%%%%%%%%%%%%%%%%%%%%%%%%%%%%%%%%%%%%%
Using these relations, the operation of $W_p$ for the spin states on the honeycomb plaquette is given by
% Then,
% the local conserved quantity Eq.~(\ref{eq:cons}) has the following matrix elements as
\begin{eqnarray}
 &&W_p \ket{m_1}_1 \ket{m_1}_2 \ket{m_3}_3 \ket{m_4}_4 \ket{m_4}_5 \ket{m_6}_6\nonumber \\
%  &= \mp (-i)^{2\left(m_1+m_3+m_4+m_6\right)} \ket{-m_1}_1 \ket{-m_1}_2 \ket{m_3}_3 \ket{-m_4}_4 \ket{-m_4}_5 \ket{m_6}_6.
  =&\mp& \left[e^{-i\pi m_1}\ket{-m_1}_1 \ket{-m_1}_2\right]\left[e^{-i\pi m_3}\ket{m_3}_3\right]\nonumber\\
  &\times&\left[e^{-i\pi m_4}\ket{-m_4}_3 \ket{-m_4}_4\right] \left[e^{-i\pi m_6}\ket{m_6}_6\right],
\end{eqnarray}
where $m_i$ ($i=1,3,4,6$) takes $\pm S$.
In the integer spin case,
the prefactor $q_i=\exp[-i\pi m_i]$ is independent of its local magnetization $m_i$
as $q_i=(-1)^S$.
On the other hand, this clearly depends on $m_i$
as $q_i=(-1)^S {\rm sgn}(m_i)$ for half-integer spins.
% Note that, when the low-energy subspace is focused on $(m_i=\pm S)$,
% the magnitude of the spins plays an important role.
% It is found that, for integer spins,
% the quantity $q=\exp[-i\pi m]$ is independent of
% the quamtum number $m$, where $q=(-1)^S$.
% On the other hand, $q=(-i)^{2m}$ for half-integer spins.
Therefore, $W_p^{\rm eff}$ obeys the following relation:
\begin{align}
 &W_p^{\rm eff}\kets{\tilde{\sigma}_1}_{{\rm left}(p)}\kets{\tilde{\sigma}_3}_{{\rm top}(p)}\kets{\tilde{\sigma}_4}_{{\rm right}(p)}\kets{\tilde{\sigma}_6}_{{\rm bottom}(p)}\notag\\
&=C
\kets{-\tilde{\sigma}_1}_{{\rm left}(p)}\kets{\tilde{\sigma}_3}_{{\rm top}(p)}\kets{-\tilde{\sigma}_4}_{{\rm right}(p)}\kets{\tilde{\sigma}_6}_{{\rm bottom}(p)},
\end{align}
where the coefficient is given by $C=-\tilde{\sigma}_1\tilde{\sigma}_3\tilde{\sigma}_4\tilde{\sigma}_6$
for half-integer spins, while $C=1$ for integer spins.
The pseudospin representation of $W_p^{\rm eff}$ is then obtained as
\begin{align}
 W_p^{\rm eff} =
  \begin{cases}
   \tilde{\sigma}_{{\rm left}(p)}^y\tilde{\sigma}_{{\rm top}(p)}^z\tilde{\sigma}_{{\rm right}(p)}^y\tilde{\sigma}_{{\rm bottom}(p)}^z & (S{\rm : half\ integer})\\
 \tilde{\sigma}_{{\rm left}(p)}^x\tilde{\sigma}_{{\rm right}(p)}^x & (S{\rm :integer})
  \end{cases}.
\label{eq:wpeff}
\end{align}
Note that, in the integer spin cases,
the local conserved quantity is represented only by two pseudospin operators
although it is defined in each plaquette.
This is because the phase factor for the spins at sites 3 and 6
disappears due to Eq. (\ref{eq:z}).
An important point is that the formulation Eq. (\ref{eq:wpeff})
depends on the spin magnitudes.
This result should suggest a qualitative difference
even in the ground-state properties,
which will be discussed in the following section.

%%%%%%%%%%%%%%%%%%%%%%%%%%%%%%%%%%%%%%%%%%%%%%%%%%%%%%%%%
\section{Effective model}
\label{sec:effective-model}
%%%%%%%%%%%%%%%%%%%%%%%%%%%%%%%%%%%%%%%%%%%%%%%%%%%%%%%%%

In this section, we consider the effective Hamiltonian of the spin-$S$ Kitaev model
in the anisotropic limit.
The low-energy model is obtained via the perturbation procedure with respect to the interactions on $x$ and $y$ bonds.
The $n$-th order effective Hamiltonian is formally given as
\begin{align}
 {\cal H}_{\rm eff}^{(n)} = PV\left(\frac{1}{E_0-{\cal H}_0}QV\right)^{n-1} P + \tilde{\cal H}_{\rm eff}^{(n)},
\label{eq:effective}
\end{align}
where $Q(=1-P)$ is the projection operator out of the low-energy subspace.
In this form, the first term represents contributions from the perturbation processes,
where any intermediate states do not belong to the low-energy subspace.
On the other hand, the second term $\tilde{\cal H}_{\rm eff}^{(n)}$
represents the contributions from the other processes.
In the perturbation expansions for the spin-$S$ Kitaev model, $\tilde{\cal H}_{\rm eff}^{(n)}$ merely gives a constant in the lowest-order relevant contributions.
Therefore, we focus on the first term of Eq.~(\ref{eq:effective}) to obtain the form of the effective Hamiltonian.

\begin{figure}[t]
\begin{center}
\includegraphics[width=3cm]{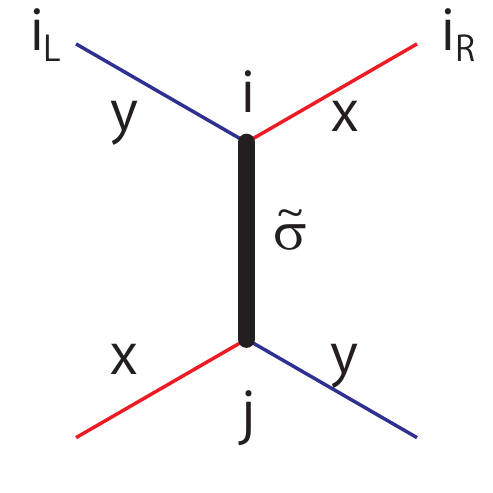}
\caption{
  Schematic picture for the perturbation bonds
  connected with a certain $z$-bond with the pseudospin $\tilde{\sigma}$.
}
\label{fig:perturb}
\end{center}
\end{figure}

Before we proceed our discussions,
we consider the key process of the perturbations.
The perturbed Hamiltonian $V$ is composed of the $x$-bond interaction $-J_xS_i^xS_j^x$ and
$y$-bond interaction $-J_y S_i^yS_j^y$.
A certain site $i$ is connected by the $x$- and/or $y$-bonds,
as shown in Fig.~\ref{fig:perturb}.
Since $x$- and $y$-component of the spin operators are
given as
$S^x=\left(S^++S^-\right)/2, S^y=\left(S^+-S^-\right)/2i$,
each bond increments or decrements the local quantum number $m$ by 1.
This yields two constraints in the possible perturbation processes.

(i) When the initial spin state at site $i$ coincides with the final one in the perturbation calculation,
the number of perturbation bonds connecting to the site $i$ must be even.
Note that, when the number of the perturbations on $y$-bonds among them is odd, the phase factor appears depending on the initial spin state $|\pm S\rangle$.
In the case where this condition is satisfied at either site $i$ or $j$ on a $z$-bond dimer $r$, the $z$-component of the pseudospin operator, $\tilde{\sigma}_r^z$, appears in the effective Hamiltonian.
Furthermore, using the above conditions, it is shown that the interaction terms consisting of the product only of $\tilde{\sigma}^z$ do not appear in the effective Hamiltonian with finite-order perturbations.
Thus, there must be pseudospins flipped in the perturbation processes.
This consideration is also supported from the viewpoint of the presence of the local conserved quantity.
In fact, the finite products $\prod_r \tilde{\sigma}_r^z$ on linked $z$ bond dimers do not commute with all local conserved quantities given in Eq.~(\ref{eq:wpeff}) in the low-energy subspace.

 (ii) $2S$-times perturbations on the bonds connecting to the site $i$ are, at least,
 needed to flip its spin state.
This means that the effective Hamiltonian should be described by the $4S$th perturbations
since $2S$-times perturbation interactions are needed for both sites $i$ and $j$ on a $z$-bond
to flip the pseudospin.
We here note that, in the half-integer case,
when $2S$-times perturbation interactions applied for the bonds
connecting to the site $i$,
either $i_R$ or $i_L$ shown in Fig.~\ref{fig:perturb} is connected by the odd number of perturbation interactions.
In the case, the constraint (i) is not satisfied in the site.
Therefore, in the effective Hamiltonian,
such perturbation processes do not contribute to the effective Hamiltonian within the $4S$th order perturbations
and $8S$th order perturbations should dominate the low-energy effective Hamiltonian.
On the other hand, in the integer spin case,
$2S$ is even, and thereby
low-energy Hamiltonian should be described by the $4S$th order perturbations.
In the following, taking into account the qualitative difference in spins,
we examine the effective Hamiltonians with $S>1/2$.

%%%%%%%%%%%%%%%%%%%%%%%%%%%%%%%%%%%%%%%%%%%%%%%%
%%%%%%%%%%%%%%%%%%%%%%%%%%%%%%%%%%%%%%%%%%%%%%%%

%%%%%%%%%%%%%%%%%%%%%%%%%%%%%%%%%%%%%%%%%%%%%%%%
\begin{figure}[t]
\begin{center}
\includegraphics[width=8cm]{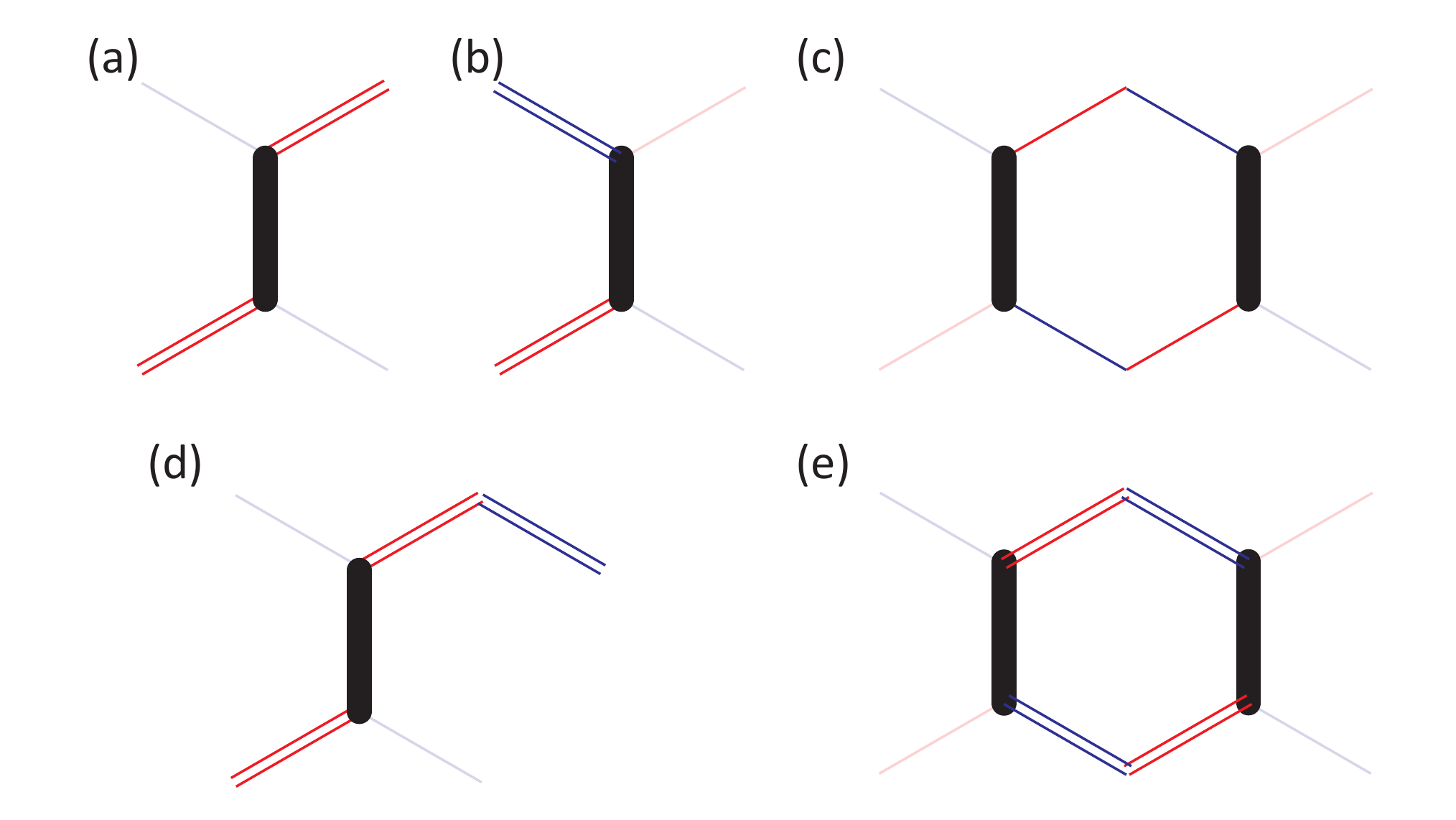}
\caption{
  Schematic pictures for the perturbation process for the $S=1$ Kitaev model.
  $x$ and $y$ perturbation bonds are expressed by the thin red and blue lines, and $z$ bonds
  are by the bond lines.
  $n$-fold $x$-bonds ($y$-bonds) mean that $n$-times $J_xS_i^xS_j^x$ ($J_yS_i^yS_j^y$)
  are considered in the perturbation process.
}
\label{fig:S1}
\end{center}
\end{figure}
%%%%%%%%%%%%%%%%%%%%%%%%%%%%%%%%%%%%%%%%%%%%%%%%

%%%%%%%%%%%%%%%%%%%%%%%%%%%
%\subsection{Integer spins}
%%%%%%%%%%%%%%%%%%%%%%%%%%%
Let us start with the $S=1$ case as a first example for integer spins.
There are four kinds of the perturbation processes relevant for the effective Hamiltonian.
Two processes are schematically depicted in Figs.~\ref{fig:S1}(a) and~\ref{fig:S1}(b).
Neighboring spins on the $z$-bond are connected by two perturbation bonds,
which flip the pseudospin defined on the $z$ bond.
On the other hand, the perturbation process with four distinct bonds,
as shown in Fig.~\ref{fig:S1}(c),
never contributes to the effective Hamiltonian
since each site on the $z$-bonds do not satisfy the constraint (ii).
This is contrast to the case of $S=1/2$ Kitaev model,
where such a process plays an essential role in the effective Hamiltonian.
By taking into account relevant perturbation processes for the $S=1$ Kitaev model,
the low-energy effective Hamiltonian is then given as
%%%%%%%%%%%%%%%%%%%%%%%%%%%%%%%%%%%%%%%%%%%%%%%%%
\begin{align}
 {\cal H}_{\rm eff}^{S=1} = -J_{\rm eff}^{S=1}\sum_{r}\tilde{\sigma}_r^x,
\label{eq:S=1,1}
\end{align}
%%%%%%%%%%%%%%%%%%%%%%%%%%%%%%%%%%%%%%%%%%%%%%%%%
where $J_{\rm eff}^{S=1}=7(J_x^2-J_y^2)^2/192J_z^3$.
This model is equivalent to an isolated spin model under the magnetic field
in $x$ direction and is not given by the sum of $W_p^{\rm eff}$ unlike the $S=1/2$ case.
Nevertheless, it is obvious that the effective Hamiltonian commutes with all $W_p^{\rm eff}$.
We wish to note that, in the case $|J_x|=|J_y|$,
$J_{\rm eff}^{S=1}$ vanishes from its explicit form.
  In the case, the effective Hamiltonian is instead
  obtained from higher-order perturbations; their examples are schematically
  shown in Figs.~\ref{fig:S1}(d) and~\ref{fig:S1}(e).
  Nevertheless, the sixth-order contributions are shown to vanish because of the same reason as the fourth-order one in the case with $|J_x|=|J_y|$.
  By performing the perturbation expansion, we obtain the effective model as follows:
  \begin{align}
    {\cal H}_{\rm eff}^{S=1} = -\tilde{J}
    \sum_p \tilde{\sigma}_{{\rm left}(p)}^x\tilde{\sigma}_{{\rm right}(p)}^x
    +\tilde{h}\sum_r \tilde{\sigma}_{\rm r}^x,\label{eq:2}
  \end{align}
  where $\tilde{J}$ and $\tilde{h}$ are effective exchange coupling and field,
  which are given by the eighth order of $|J_x|=|J_y|$.
  This Hamiltonian is regarded as longitudinal field Ising spin chains.

%\blue{In the case, the effective Hamiltonian is instead
%given by the eighth order perturbations as,
%\begin{align}
%  {\cal H}_{\rm eff}^{S=1} = -\frac{10823J_x^8}{35389440J_z^7}
%  \sum_p \tilde{\sigma}_{{\rm left}(p)}^x\tilde{\sigma}_{{\rm right}(p)}^x.
%\label{eq:S=1,2}
%\end{align}
%Namely, the sixth-order perturbations should vanish, which will be discussed.
%This effective Hamiltonian can be regarded as
%disconnected one-dimensional ferromagnetic Ising spin chains
%perpendicular to the $z$-bonds.　$8$次で縦磁場イジング模型になると本文中で述べるだけで十分な気がします。}

By performing similar calculations for $S=2$ and $3$ cases,
we obtain the effective Hamiltonians as,
\begin{eqnarray}
  {\cal H}_{\rm eff}^S &=&-J_{\rm eff}^{S}\sum_r  \tilde{\sigma}_r^x,
\end{eqnarray}
where
\begin{widetext}
  \begin{align}
    J_{\rm eff}^{S=2}=&\frac{18604521\left(J_x^8 + J_y^8\right)-82758048\left(J_x^6 J_y^2 + J_x^2 J_y^6\right) +129273554 J_x^4 J_y^4 }{5138022400J_z^7 },\\
    J_{\rm eff}^{S=3}=&\frac{\left(J_x^2-J_y^2\right)^2}
    {152769160756403896320000J_z^{11}}\nonumber\\
&\times\left[36052814083126422740 \left(J_x^8 + J_y^8\right)
 -176028114277347622010 \left(J_x^6 J_y^2 + J_x^2 J_y^6\right)
 + 287126525350219384887 J_x^4 J_y^4\right].
\end{align}
\end{widetext}
The system can be regarded as isolated spins under the magnetic field, which is essentially the same
as the $S=1$ system.
Therefore, in the integer spin Kitaev models,
quantum fluctuations are quenched due to the anisotropy of interactions
and the system becomes classical in the limit.
Then, the ground state is fully polarized for the $\tilde{\sigma}^x$ direction, and $W_p=1$.
Its excitation energy is given by $\Delta=2J_{\rm eff}^S$.
Note that, in the $S=2$ case,
the effective Hamiltonian never disappears even in the $|J_x|=|J_y|$ case,
which is contrast to the $S=1$ and $S=3$ cases.

%%%%%%%%%%%%%%%%%%%%%%%%%%%%%%%%%%%%%%%%%%%%%%%
\begin{figure}[t]
 \begin{center}
  \includegraphics[width=\columnwidth,clip]{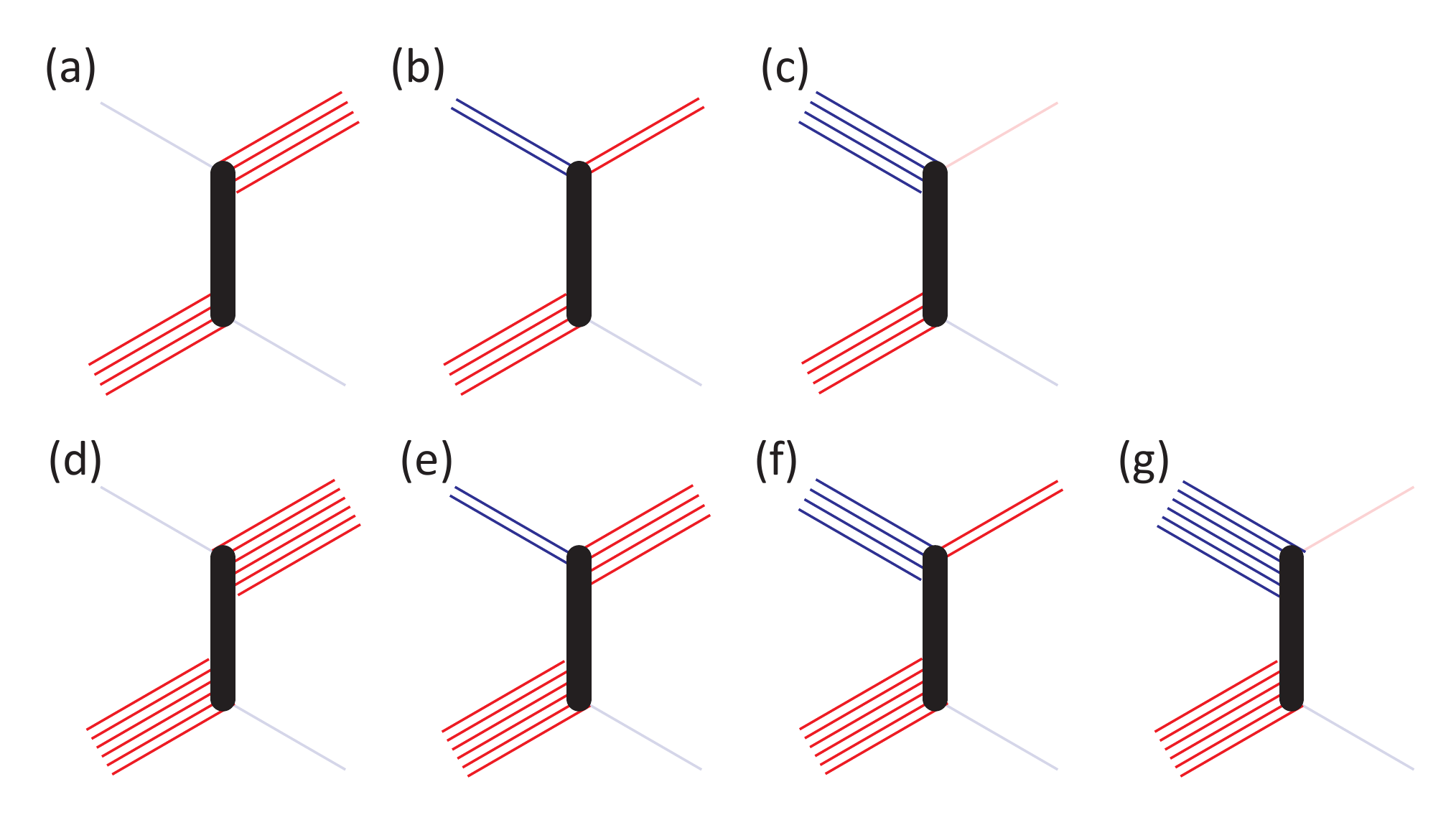}
  \caption{
    Examples of perturbation processes in the integer spin Kitaev models.
    (a)-(c) for $S=2$ and (d)-(g) for $S=3$.
}
  \label{fig:SS}
 \end{center}
\end{figure}
%%%%%%%%%%%%%%%%%%%%%%%%%%%%%%%%%%%%%%%%%%%%%%%

To address this issue, we focus on the phase factor for
the perturbation processes in the integer spin cases.
Figures~\ref{fig:S1} (a) and~\ref{fig:S1}(b) show two of relevant perturbation processes
in the $S=1$ case.
In these perturbation processes,
a local spin state for a certain site flips due to two $x$-bond interactions
[Fig.~\ref{fig:S1}(a)] and $y$-bond interactions [Fig.~\ref{fig:S1}(b)],
which lead to the contributions with different signs, $J_x^2$ and $-J_y^2$, respectively.
Therefore, the contribution is canceled out in the case $|J_x|=|J_y|$.
On the other hand, different behavior appears in the $S=2$ case.
Figures~\ref{fig:SS}(a)--\ref{fig:SS}(c) show the perturbation processes
for the $S=2$ case.
The corresponding factors are generally given as $J_x^4$, $-J_x^2J_y^2$, and $J_y^4$,
which should lead to finite contributions.
Some relevant perturbation processes for the $S=3$ case are shown in Figs.~\ref{fig:SS}(d)--\ref{fig:SS}(g).
According to the above discussions, these perturbation processes yield
$J_x^6$, $-J_x^4J_y^2$, $J_x^2J_y^4$, and $-J_y^6$, respectively.
Therefore, they are exactly cancelled out in the case $|J_x|=|J_y|$.
In the odd spin $S$ case with $|J_x|=|J_y|$, we have confirmed that
the effective Hamiltonian are generally described by Eq.~(\ref{eq:2}), where $\tilde{J}$ ($\tilde{h}$) are given by $8S$th [$(4S+4)$th] order of $|J_x|=|J_y|$.

%%%%%%%%%%%%%%%%%%%%%%%%%%%%%%%%%%%%%%%%%%%%%%%
\begin{figure}[t]
 \begin{center}
  \includegraphics[width=\columnwidth,clip]{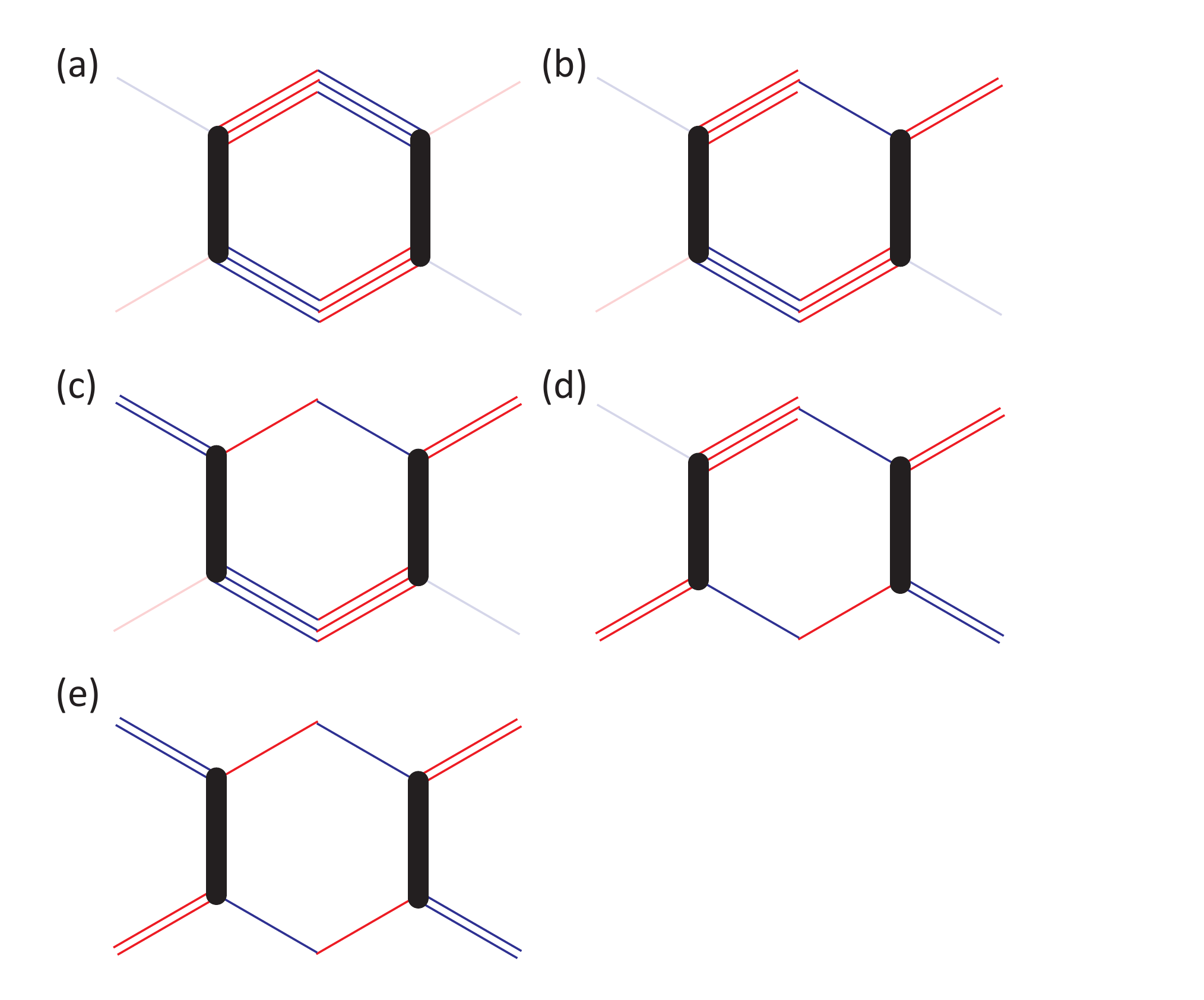}
  \caption{
    Example of sixth-order perturbation processes in the $S=3/2$ Kitaev model.
}
  \label{fig:picture13}
 \end{center}
\end{figure}

%%%%%%%%%%%%%%%%%%%%%%%%%%%%%%%%%%%%%%%%%%%%%%%
Next, we consider the $S=3/2$ spin model, as an example for half-integer spins.
% Due to two constraints in the perturbation processes,
% there are no relevant $4S$th order perturbations in the effective Hamiltonian, as mentioned.
As mentioned before, the lowest order contribution is of $8S$th order.
Five of these processes are schematically shown in Fig.~\ref{fig:picture13}.
First, we consider the perturbation process shown in Fig.~\ref{fig:picture13}(a).
In this process, two pseudospins on each $z$-bond are flipped since three perturbations are applied to the $x$- and/or $y$-bonds connected to them.
On the other hand, the top and bottom sites of the hexagon in Fig.~\ref{fig:picture13} contribute as $\tilde{\sigma}^z$ in the effective Hamiltonian due to the constraint (i).
Therefore, the process in Fig.~\ref{fig:picture13}(a) is described by
$\tilde{\sigma}_{{\rm left}(p)}^y\tilde{\sigma}_{{\rm top}(p)}^z\tilde{\sigma}_{{\rm right}(p)}^y\tilde{\sigma}_{{\rm bottom}(p)}^z$.
Other possible perturbation processes satisfying two constraints also
have a loop structure in the bond configuration, as shown in Fig.~\ref{fig:picture13}.
This is contrast to that for the integer spin cases.
By taking into account all possible perturbation processes,
we obtain the effective Hamiltonian
%%%%%%%%%%%%%%%%%%%%%%%%%%%%%%%%%%%%%%%%%
\begin{widetext}
\begin{align}
 {\cal H}_{\rm eff}^{S=3/2} = -&\left[3214648723397092084  J_x^6 J_y^6 +1646995686930432837306 J_x^2 J_y^2 \left(J_x^2 -J_y^2\right)^4  + 91522768044989658195 J_x^4 J_y^4 \left(J_x^2 -J_y^2\right)^2\right] \nonumber\\
&\times\frac{1}{3076979551468152422400000 J_z^{11}} \sum_p \tilde{\sigma}_{{\rm left}(p)}^y\tilde{\sigma}_{{\rm top}(p)}^z\tilde{\sigma}_{{\rm right}(p)}^y\tilde{\sigma}_{{\rm bottom}(p)}^z.
\end{align}
\end{widetext}
%%%%%%%%%%%%%%%%%%%%%%%%%%%%%%%%%%%%%%%%%

As seen above, we can expect that there exists a loop structure for relevant perturbation processes
in half-integer spin Kitaev models.
In the case, the effective Hamiltonian in general should be given by
\begin{eqnarray}
  {\cal H}_{\rm eff}^{S} &=&-J_{{\rm eff}}^{S}\displaystyle{\sum_p} \tilde{\sigma}_{{\rm left}(p)}^y\tilde{\sigma}_{{\rm top}(p)}^z\tilde{\sigma}_{{\rm right}(p)}^y\tilde{\sigma}_{{\rm bottom}(p)}^z\\
  &=&-J_{{\rm eff}}^{S}\displaystyle{\sum_p}W_p^{\rm eff},\label{eq:1}
\end{eqnarray}
where the coupling constant $J_{{\rm eff}}^{S}$ is given in the $8S$th order with respect to $J_x$ and/or $J_y$.
This model is essentially the same as the low-energy Hamiltonian for $S=1/2$,
and quantum fluctuations play a crucial role in the ground state.
The low-energy excitations are described by the anyons and
the ground state exhibits nontrivial degeneracy due to the topological order.

\begin{figure}[t]
 \begin{center}
  \includegraphics[width=0.9\columnwidth]{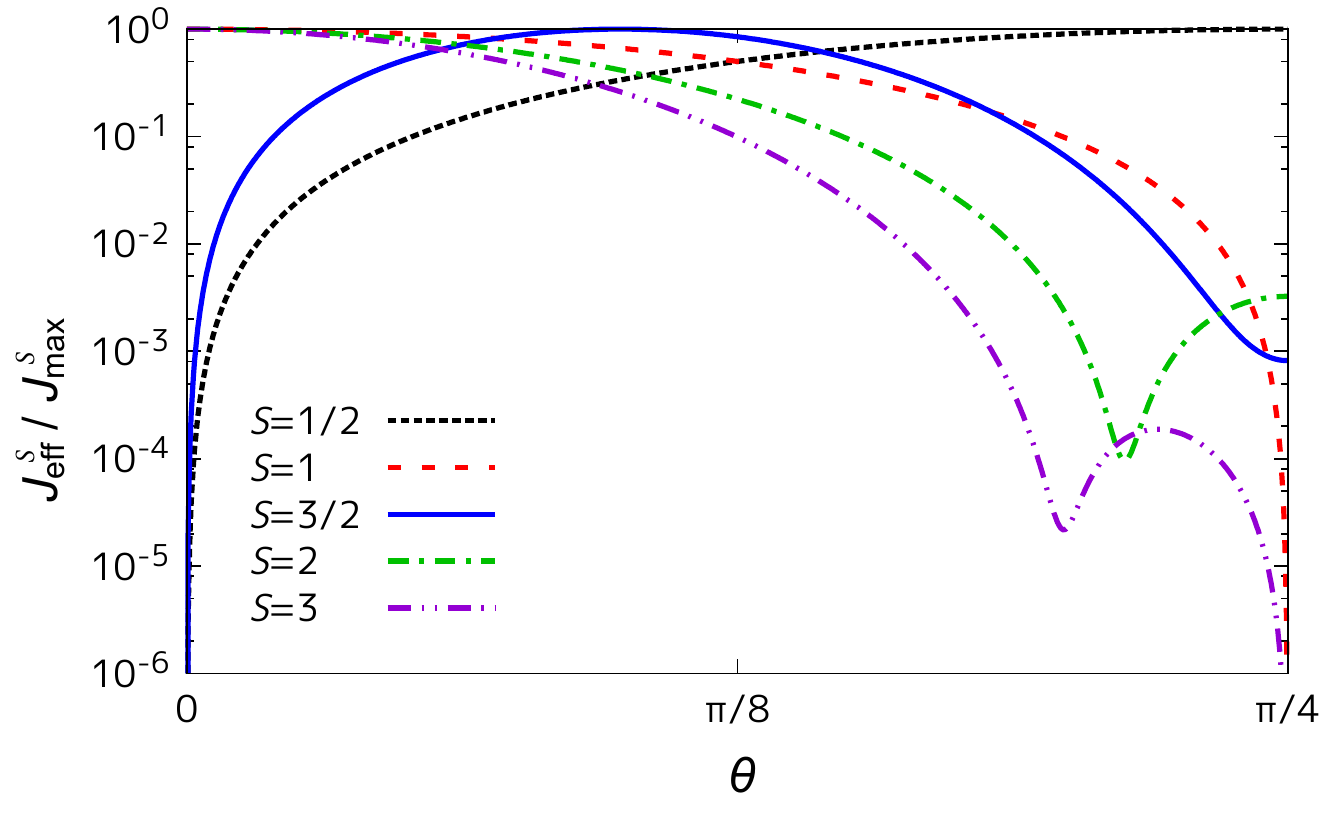}
  \caption{
  Coefficient of the effective Hamiltonian, $J_{{\rm eff}}^S$, as a function of $\theta=\arctan(J_y/J_x)$.
  Here, $J_{{\rm eff}}^S$ is normalized by its maximum $J_{{\rm max}}^S$, and $J_{{\rm eff}}^S/J_{{\rm max}}^S$ does not depend on $J_z$.
  }
  \label{fig:coefficient}
 \end{center}
\end{figure}

We here discuss how the imbalance between the exchange constants $J_x$ and $J_y$ affects
the coupling constant of the effective Hamiltonians.
We introduce the parameter $\theta$ so that $J_x=J\cos\theta$ and $J_y=J\sin\theta$, and examine
the coupling constant $J_{{\rm eff}}^{S}$ normalized by its maximum,
as shown in Fig.~\ref{fig:coefficient}.
First, we focus on the integer spin Kitaev models.
We find that the coefficients have a maximum at $\theta=0$ and
are finite except for the case with $\theta=\pi/4$ and odd $S$.
In addition, nonmonotonic behavior appears in the integer spin system with $S>1$,
and the coupling constant has a minimum around $\theta=0.669$ ($0.626$)
for the $S=2$ ($3$) system.
As for the half-integer spin cases, monotonic behavior appears in the $S=1/2$ case, while
nonmonotonic behavior in the $S=3/2$ case.
%%%%%%%%%%%%%%%%%%%%%%%%%%%%%%%%%%%%%%%%%%%%%%%%%%
%%%%%%%%%%%%%%%%%%%%%%%%%%%%%%%%%%%%%%%%%%%%%%%%%%%
It is expected that the coupling constants are also positive in the other $S$ cases,
which is consistent with the previous theoretical work~\cite{PhysRevB.78.115116},
where $W_p$ is always unity in the ground state
in the original spin-$S$ Kitaev models.

%%%%%%%%%%%%%%%%%%%%%%%%%%%%%%%%%%%%%%%%%%%%%%%%%%%
%\section{Boundary effect}
%\label{sec:boundary-effect}
%%%%%%%%%%%%%%%%%%%%%%%%%%%%%%%%%%%%%%%%%%%%%%%%%%%

\begin{figure}[t]
 \begin{center}
  \includegraphics[width=\columnwidth,clip]{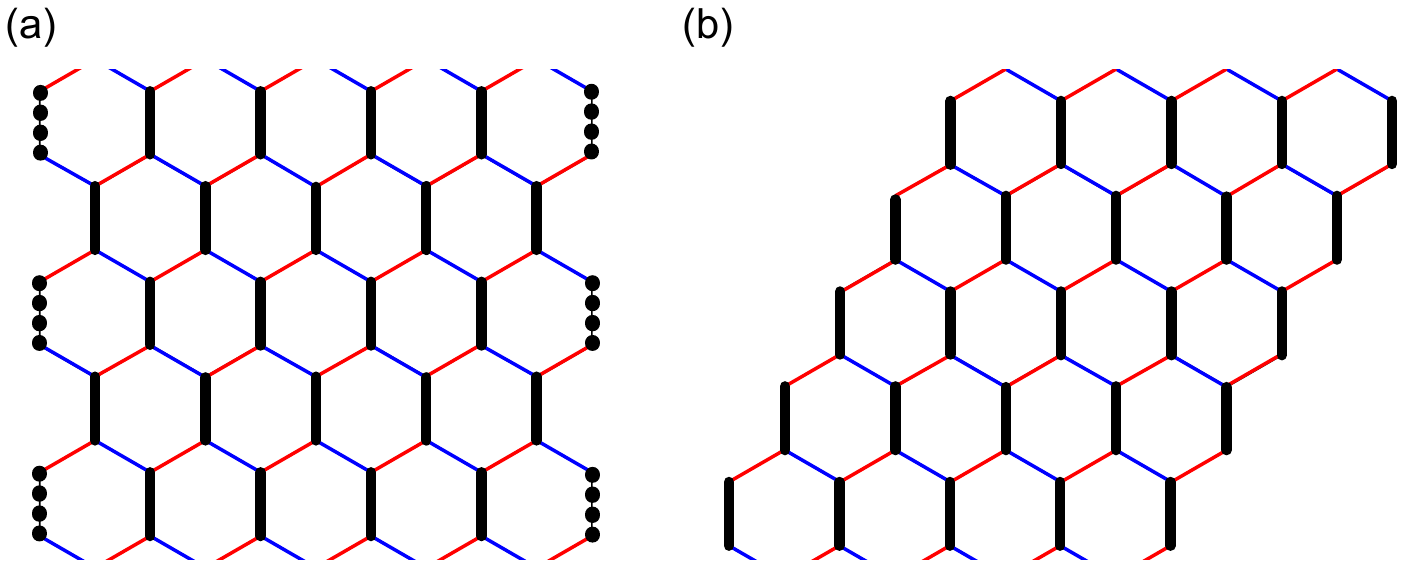}
  \caption{
    Lattice structures associated with (a) the armchair-type and
    (b) zigzag-type edges.
On the dotted bonds shown in (a), the relevant contributions given in Eq.~(\ref{eq:edge}) appears.
  }
  \label{fig:edge}
 \end{center}
\end{figure}

Before closing the section, we would like to discuss how the boundary of the lattice
affects ground-state properties in the Kitaev model.
Here, we consider armchair and zigzag type edges, as shown in Fig.~\ref{fig:edge}.
%%%%%%%%%%%%%%%%%%%%%%%%%%%%%%%%%%%%%%%%%%%%
%%%%%%%%%%%%%%%%%%%%%%%%%%%%%%%%%%%%%%%%%%%%
As discussed in the previous section,
we have considered the effective Hamiltonian via the virtual processes.
Open boundary conditions may rule out some perturbation processes,
leading to peculiar boundary effects in the effective Hamiltonian.
In particular, boundary contributions play an essential role for the odd-spin systems with $|J_x|=|J_y|$.
When the system has the armchair type edge,
the lack of some perturbation process induces the $4S$th order perturbations
for the pseudospins on the edge.
In fact, $z$-bonds on the right edge in Fig.~\ref{fig:edge}(a),
the perturbation process shown in Fig.~\ref{fig:S1}(a) is ruled out.
Therefore, in the effective Hamiltonian,
the dominant $4S$th perturbations appear around the edges,
while the effective Hamiltonian in the bulk is given by higher order perturbations.
In this case, the lowest-order effective Hamiltonian is given as
\begin{eqnarray}
  {\cal H}_{\rm eff}&=&J_{\rm edge}\sum_{r\in \rm edge} \tilde{\sigma}_r^x,
\label{eq:edge}
\end{eqnarray}
where $J_{\rm edge}\sim O(J_x^{2S}J_y^{2S})$.
On the other hand, around the boundary with the zigzag structure,
the above boundary term does not appear because of the following reason.
For example, if we focus on a right edge in the $S=1$ Kitaev model
[see Fig.~\ref{fig:edge}(b)],
two perturbation processes shown in Fig.~\ref{fig:S1}(a) and~\ref{fig:S1}(b) are allowed and
these contributions are cancelled out in the $|J_x|=|J_y|$ case.
Therefore, such an open boundary leads to no drastic change
in the Kitaev model with zigzag edges.

%%%%%%%%%%%%%%%%%%%%%%%%%%%%%%%%%%%%%%%
\section{Summary}
\label{sec:summary}
%%%%%%%%%%%%%%%%%%%%%%%%%%%%%%%%%%%%%%%

We have investigated low-energy properties of the anisotropic limit of the spin-$S$ Kitaev model.
We have obtained low-energy representation of a local conserved quantity,
which should play a key role in stabilizing the quantum spin liquid state.
It has been found that the effective form of the local conserved quantity for the half-integer spins is
different from that for the integer spins.
Applying the perturbation theory to the anisotropic Kitaev model,
we have obtained the effective Hamiltonian.
In the integer spin case,
the effective model is given by a noninteracting spin model under an uniform magnetic field,
which is a classical system without quantum fluctuation.
On the other hand, in the half-integer case,
the system is described by the Hamiltonian of the toric code,
where quantum fluctuations in pseudospins play a crucial role in the ground state.

The effective Hamiltonian is given by the $4S$th ($8S$th) order perturbation expansion
in the integer (half-integer) spin Kitaev model in the anisotropic limit.
Therefore, its energy scale is much lower than the exchange coupling $J_z$.
This means that low-energy physics described by the effective Hamiltonian
appears at very low temperatures.
This is consistent with the numerical results
for the anisotropic Kitaev model~\cite{oitmaa2018incipient},
where the clear plateau at $S=\frac{1}{2}\ln 2$ appears in the entropy of the anisotropic system.

We have clarified how the qualitative difference between integer and half-integer spins appears
in the effective Hamiltonian describing low-energy properties.
The spin dependence in isotropic spin-$S$ Kitaev models and their implementation in real materials remain interesting problems, which are now under consideration.

\begin{acknowledgments}
%  We thank ***.
  This work was supported by Grant-in-Aid for Scientific Research from
  JSPS, KAKENHI Grant Nos. JP18K04678, JP17K05536 (A.K.),
  JP16K17747, JP16H02206, JP18H04223 (J.N.).
%  Parts of the numerical calculations are performed
%  in the supercomputing systems in ISSP, the University of Tokyo.
\end{acknowledgments}

\bibliography{./refs}

\end{document}